\documentclass[%
 reprint,
superscriptaddress,
nofootinbib,
 amsmath,
 aps,
 prl,
floatfix,
]{revtex4-2}

\usepackage{graphicx}
\usepackage{dcolumn}
\usepackage{bm}

\usepackage{enumitem}
\usepackage{epsfig}
\usepackage{bm}
\usepackage{amsmath}
\input{epsf}
\usepackage{psfrag}
\usepackage{subcaption}
\usepackage{xcolor}
\usepackage{pdfpages}
\usepackage[normalem]{ulem}
\makeatletter
\AtBeginDocument{\let\LS@rot\@undefined}
\makeatother
\newcommand{\bea}{\begin{eqnarray}}
\newcommand{\eea}{\end{eqnarray}}

\newcommand{\nn}{\nonumber}

\newif\ifshowcomments
\showcommentstrue 

\begin{document}

\title{Sivers Tomography from Charge and Angle Only}

\author{Haotian Cao}
\email{haotiao.cao@northwestern.edu}
 \affiliation{Department of Physics \& Astronomy, Northwestern University, Evanston, Illinois 60208, USA}
\affiliation{Joint BNL-SBU Center for Frontiers in Nuclear Science (CFNS),
Stony Brook University, Stony Brook, New York 11794, USA}

\author{Xiaohui Liu}
 \email{xiliu@bnu.edu.cn}
 \affiliation{School of Physics and Astronomy, Beijing Normal University, and Key Laboratory of Multiscale Spin Physics (Beijing Normal University), Ministry of Education, Beijing 100875, China}
 \affiliation{Southern Center for Nuclear Science Theory (SCNT), Institute of Modern Physics, Chinese Academy of Science, Huizhou 516000, China}

\author{Frank~Petriello}
\email{f-petriello@northwestern.edu}
 \affiliation{Department of Physics \& Astronomy, Northwestern University, Evanston, Illinois 60208, USA}

\begin{abstract}
We propose a one-point charge-correlator (OPCC) probe of the Sivers effect in back-to-back deep-inelastic scattering. This measurement uses only the signs and directions of charged tracks, with no calorimetric or particle-identification information required. The observable weights the final state by its electric charge and measures the azimuthal correlation between the charge flow and the transverse spin of the proton. This probe is shown to be IRC finite and admits a factorization involving the usual Sivers distribution and a perturbatively calculable charge-weighted jet function for small transverse seperation $b\ll \Lambda_{\rm QCD}^{-1}$, with no reliance on non-perturbative fragmentation functions or track functions due to charge conservation. We validate the factorization against the full fixed-order QCD and present resummed predictions at N\(^3\)LL accuracy for the unpolarized distribution and N\(^2\)LL for the Sivers asymmetry. The OPCC provides a theoretically clean and simple experimental measurement, and establishes a charge-and-angle measurement paradigm for spin physics at a future Electron-Ion Collider. 
\end{abstract}

\maketitle
 
\textbf{\emph{{\color{magenta}Introduction.}}}
The spin structure of the nucleon remains one of the most direct windows into the non-perturbative dynamics of QCD~\cite{AbdulKhalek:2021gbh,Burkert:2022hjz,Boussarie:2023izj}. Beyond the decomposition of the proton spin into quark and gluon helicities, transverse-spin observables expose a more differential form of nucleon structure~\cite{Boussarie:2023izj}. The Sivers function is a central example of how spin reveals novel aspects of nucleon structure~\cite{Sivers:1989cc}. It correlates the transverse spin of the proton with the intrinsic transverse motion of an unpolarized parton, and therefore probes the spin-orbit and gauge-link dynamics that are invisible in collinear spin measurements~\cite{Brodsky:2002cx,Ji:2002aa,Belitsky:2002sm}.

Energy correlators~\cite{Basham:1978zq,Basham:1978bw,Basham:1977iq,Ore:1979ry,Sveshnikov:1995vi,Korchemsky:1997sy,Korchemsky:1999kt,Belitsky:2001ij, Lee:2006nr,Hofman:2008ar,Moult:2025nhu} have recently opened a new window into this study~\cite{Gao:2025evv,Song:2025bdj,Kang:2026pro}. They measure angular patterns in the flow of final-state energy. In the back-to-back region these angular correlations are governed by the same recoil dynamics that appears in TMD factorization~\cite{Moult:2018jzp,Kang:2023big,Kang:2024dja,Bhattacharya:2025bqa,Gao:2025cwy}, while in the target fragmentation region they provide complementary access to spin-dependent parton motion~\cite{Liu:2022wop,Cao:2023oef,Liu:2024kqt,Chen:2024bpj,Zhu:2025qkx,Gao:2025cwy}.

There is another class of angular correlators that has been much less
studied in spin physics.  We can replace the energy-flow operator by the flux of a
conserved current.  For electric charge, the corresponding detector operator is~\cite{Hofman:2008ar,Chicherin:2020azt,Cuomo:2025pjp,Riembau:2024tom,Monni:2025zyv}
\begin{align} \label{eq:Q-detector}
\mathcal Q(\vec n) = \lim_{r\to\infty} r^2\, \int_{-\infty}^\infty dt \,  n^i\, \hat{J}_{i }(t,r\vec n)\,,
\end{align}
where $\hat{J}_i$ is the electromagnetic charge current operator. The charge-detector operator $\mathcal Q(\vec n)$ measures the net electric charge
flowing in the direction \(\vec n = (\theta,\phi)\)
\begin{align} 
\label{eq:charge_flow_eigenvalue}
{\cal Q}(\vec n)\,|X\rangle
=
\sum_{h\in X} Q_h\,
\delta^{(2)}(\vec n-\vec n_h)\,
|X\rangle \,.
\end{align}
For the one-point charge correlator (OPCC) in the transversely-polarized DIS process, we measure the charge-weighted cross section such that 
\begin{align} \label{eq:OPCC-w}
\Sigma_{{\cal Q}}(\vec n,x_B,Q^2) = \sum_h \int d\sigma_{ep^\uparrow \to e h + X} \, Q_h \, \delta(\vec n - \vec n_h) \,.
\end{align} 
Equivalently, the observable can be defined by a single insertion of the charge detector operator
\begin{align} \label{eq:OPCC-o}
\Sigma_{{\cal Q}}(\vec n,x_B,Q^2) = & \frac{\alpha^2}{Q^4} 
\sum_{f,\lambda=L,T} e_f^2 f_\lambda e^\ast _{\lambda\mu}e_{\lambda\nu}  
\nonumber \\ 
  \times  
\int d^4 x \, 
e^{iq\cdot x } & 
\langle P,S_t | J_f^\mu(x)
\mathcal Q(\vec n) J_f^{\nu} (0) |P,S_t \rangle  \,. 
\end{align}
Here, $Q^2 = - q^2$ is the virtuality of the virtual photon and $e^\mu_\lambda$, with $\lambda = L,T$, denotes its polarization. $f_T = 1-y+y^2/2$ and $f_L =2-2y$. $x_B$ is the Bjorken variable and $y = Q^2/(x_B s)$ is the inelasticity with $s$ the center of mass energy squared. 

Although electric-charge correlators are naturally track-based and benefit from excellent angular resolution without requiring jet reconstruction or a full calorimetric energy measurement, they have received much less attention than energy correlators in nucleon-spin studies. Unlike energy weights, charge weights do not vanish in the soft limit.  Generic charge-sensitive angular correlations are therefore usually not regarded as IRC-safe observables.  As a result, they lack the level of theoretical control needed for systematically improvable higher-order calculations, clean factorization, and a controlled separation of perturbative radiation from non-perturbative hadronization effects.

\begin{figure}[htbp]
\begin{center}
\includegraphics[width=0.49\textwidth]{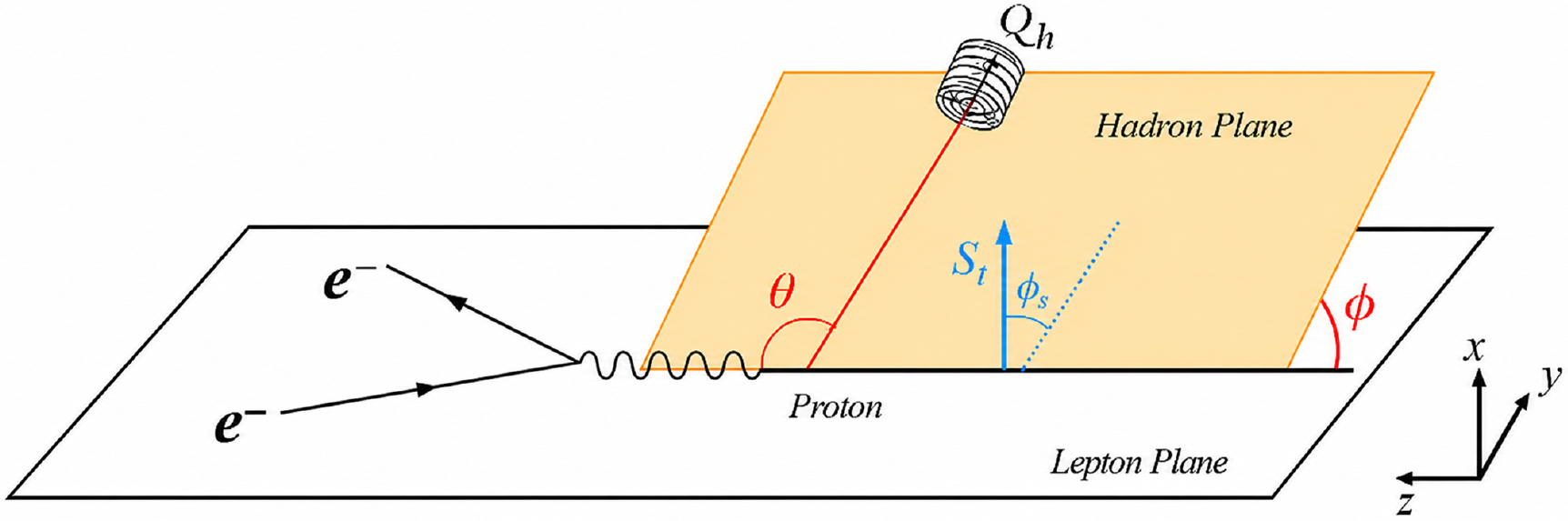}
\caption{Sketch of the OPCC observable in transversely polarized DIS showing the measured angles in the Breit frame.}
\label{fg:obs}
\end{center}
\end{figure}

The situation changes qualitatively when the OPCC in Eq.~\eqref{eq:OPCC-w} is measured, especially in a back-to-back configuration as $\theta \to \pi$ in the Breit frame
, as illustrated in Fig.~\ref{fg:obs}. Building on the analysis of the charge-charge correlator in unpolarized \(e^+e^-\) annihilation~\cite{Monni:2025zyv}, with additional inputs from charge-conjugation symmetry in the soft sector, we will show that the OPCC in back-to-back DIS admits a controlled TMD factorization and is IRC finite. This turns an experimentally advantageous measurement, based only on charged-track angles and charge signs, into an ideal probe of the Sivers effect at a future electron-ion collider (EIC) and similar facilities.

\textbf{\emph{{\color{magenta}TMDs with a One Point Charge Correlator.}}}
We consider the back-to-back region illustrated in Fig.~\ref{fg:obs}, where the charge-flow direction is measured at $(\theta,\phi)$ with $\theta \to \pi$ in the Breit frame. It is useful to introduce 
\begin{align}
\bar\theta \equiv \pi-\theta \ll 1,
\end{align}
so that \(q_T = \frac{Q\bar\theta}{2}\) is the transverse momentum of the measured charge-flow direction. For transversely-polarized proton beams, the OPCC cross section in Eq.~\eqref{eq:OPCC-o} contains the angular structures  
\begin{align}
\Sigma_{\cal Q} (\vec{n},x_B,Q^2) 
= 
 F_{UU} +  |S_t|  \sin(\phi-\phi_S)\,
F_{UT} + \dots \,.
  \label{eq:aut}
\end{align} 
The first term is the charge-weighted unpolarized structure function, while the second term is the charge-flow analogue of the Sivers modulation. The ellipses denotes other angular structure not relevant to our analysis.

Following the standard mode decomposition and field redefinition procedure in the soft-collinear effective theory (SCET) framework~\cite{Bauer:2000ew, Bauer:2000yr, Bauer:2001ct, Bauer:2001yt}, we find that at the leading power in ${\bar \theta}^2$, the unpolarized and polarized structure functions in the Breit frame admit the TMD factorization~\cite{Cao:unpol}  
\begin{align}
\label{eq:TMDunpolfac}
    F_{UU}& = \sigma_0
        H(Q^2,\mu)   \frac{{\bar \theta} Q^2}{4}  \int_0^{\infty} \frac{b\, db}{2\pi}J_0\left( b \frac{Q{\bar \theta}}{2}  \right)\nn \\
         \times & \sum_fe_f^2
         J_{f,{\cal Q}}(b,\mu,\nu) 
         f_{1}^{f} (x_B,b,\mu,\nu)S(b,\mu,\nu) \, ,
\end{align} 
and
\begin{align}
\label{eq:TMDpolfac}
    F_{UT}& = \sigma_0
        H(Q^2,\mu)  \frac{{\bar \theta}Q^2}{4}  \int_0^{\infty} \frac{b^2\, db}{4\pi}J_1\left( b \frac{Q{\bar \theta}}{2}  \right)\nn \\
         \times & \sum_fe_f^2
         J_{f,{\cal Q}}(b,\mu,\nu) 
         f_{1 T}^{\perp,f}(x_B,b,\mu,\nu)S(b,\mu,\nu) \, ,
\end{align} 
where \(J_0\) and \(J_1\) are Bessel functions of the first kind. $\sigma_0$ denotes the Born normalization, and $H$ is the hard function. The functions $f_1^f$ and $f_{1T}^{\perp,f}$ are the unsubtracted unpolarized TMD and the unsubtracted Sivers function, respectively. We can absorb the square root of the soft factor \(S\) into the TMD distributions  and the jet function $J_{f,{\cal Q}}$. 

We note that the charge detector only acts on the jet function $J_{f,{\cal Q}}$. It does not introduce a new charge-dependent soft function. The soft factor $S$ is a vacuum matrix element of Wilson lines $Y_n$/$Y_{\bar n}$ and in particular, a single insertion of the charge detector ${\cal Q}$ into the soft Wilson-line matrix element vanishes by charge conjugation, $\langle Y_nY_{\bar n}^\dagger(b) {\cal Q}(\vec n) Y_{\bar n}Y_n^\dagger(0) \rangle_\Omega = 0$, so the soft function remains the standard SIDIS TMD soft function.

The jet function appearing in Eqs.~\eqref{eq:TMDunpolfac} and~\eqref{eq:TMDpolfac} can be related to the charge-weighted TMD fragmentation function $D_{T,h\leftarrow f }$
\begin{align}
J_{f,{\cal Q}}(b)  = \sum_h \int_0^1 dz_h  Q_h D_{T,h\leftarrow f }(z_h,b) \,. 
\end{align} 
By charge conjugation, $J_{f,{\cal Q}} = - J_{{\bar f},{\cal Q}}$ and therefore the integration is well-defined when $z_h \to 0$~\cite{Monni:2025zyv,Ebert:2020qef,Luo:2020epw}. 

For perturbative transverse separations $b \ll \Lambda^{-1}_{\rm QCD}$, up to ${\cal O}(b^2 \Lambda_{\rm QCD}^2)$ corrections, the jet function can be further factorized onto the collinear fragmentation function $D_{h\leftarrow i}$, which gives~\cite{Monni:2025zyv}
\begin{align} \label{eq:jet-Q}
J_{f,{\cal Q}} 
&= \sum_{h,i}  \int dz_h Q_h\, \int_{z_h}^1 \frac{dz}{z}\,
D_{h\leftarrow i }\left(\frac{z_h}{z}\right)\,
{\cal I}_{i\leftarrow f}\left(z,b\right)   \nonumber \\ 
& = \sum_{h,i}   \int d z_h  Q_h\, 
D_{h \leftarrow i } ( z_h ) 
\int dz \, 
{\cal I}_{i\leftarrow f}(z,b)  \nonumber \\ 
& = \sum_i \,  \int dz \, Q_i \, 
{\cal I}_{i\leftarrow f}(z,b)    \,,
\end{align} 
where charge conservation is used in the fragmentation process, such that $Q_i = \sum_h  \int dz_h \, Q_h D_{h\leftarrow i} (z_h)$. 
~\footnote{Possible \(z\simeq0\) caveats associated with the 
  standard charge-conservation sum rule for the charge-odd collinear-fragmentation combination assumed here have been discussed in~\cite{Collins:2023cuo,Kotlorz:2025xso}. For the OPCC such a soft \(z\approx 0\) contribution can be assigned to the soft/zero-bin sector, whose one-point charge insertion vanishes by charge-conjugation symmetry.}. From Eq.~\eqref{eq:jet-Q}, we can see clearly that the jet function is perturbatively calculable and contains no independent non-perturbative fragmentation function
or track function. 

Therefore, in the back-to-back TMD region, the apparent difficulty of a charge-weighted final state is replaced by a standard TMD factorization formula with a perturbatively-calculable charge-weighted jet function. Currently, all ingredients in the factorization formula are under good theoretical control. The hard function $H$ is known to $4$-loops~\cite{Lee:2022nhh,Chakraborty:2022yan}, the matching of the unpolarized TMD, the jet function, and the soft function are all known to $3$-loops~\cite{Luo:2020epw,vita_2025_16658462,Li:2016ctv}. The Sivers distribution matches at small \(b\) onto the twist-three collinear quark-gluon correlation function, known as the Qiu-Sterman function~\cite{Qiu:1991pp}, $T_{F\, i/p}({x}_1,{x}_2,\mu)$

\begin{align}\label{eq:SivmQS}
   & f_{1T}^{\perp,f} (x, b,\mu,\nu) \nonumber \\ 
    = & \sum_i \int_{x}^{1} \frac{d{x}_1}{{x}_1}\frac{d {x}_2}{{x}_2} C_{f\leftarrow i}\left(\frac{x}{{x}_1},\frac{x}{{x}_2},b;\mu,\nu\right) \, 
T_{F\, i/p}({x}_1,{x}_2,\mu) \,.
\end{align}
 The matching accuracy is less advanced than the unpolarized case, but is still known to NLO~\cite{Sun:2013hua,Dai:2014ala,Scimemi:2019gge}. The TMD resummation can then be achieved at N$^3$LL and N$^2$LL accuracy for $F_{UU}$ and $F_{UT}$, respectively, with available theory inputs. The OPCC turns charge flow into a precision handle on the Sivers
effect.  It reduces the TMD measurement to track directions and charge signs
alone, avoiding the major experimental uncertainties associated with hadron identification, jet reconstruction, and calorimetry. Its factorized description can be developed with high theoretical precision.

\begin{figure}[!htbp]
  \centering
    \includegraphics[width=0.47\textwidth]{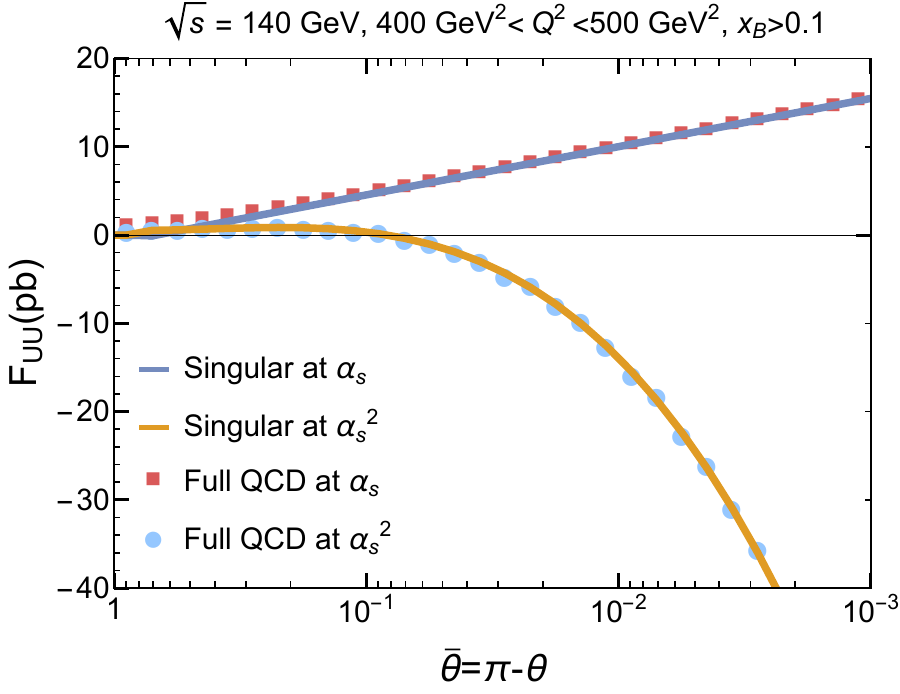}
    \caption{Comparison of the $\ln\bar \theta$ leading singular contributions from the factorization theorem 
with the full fixed-order calculations.}
  \label{fg:sing}
\end{figure}
\textbf{\emph{{\color{magenta}Validation of the Factorization.}}}
We first validate our factorization formula using the unpolarized structure function $F_{UU}$. Fig.~\ref{fg:sing} compares the full QCD calculation of the OPCC in Eq.~\eqref{eq:OPCC-w}, obtained with {\tt distress}~\cite{Abelof:2016pby} with the singular contributions predicted by the factorization theorem in Eq.~\eqref{eq:TMDunpolfac} at small $\bar \theta$. The two calculations agree very well in the
back-to-back region, both at LO (\({\cal O}(\alpha_s)\)) and NLO (\({\cal O}(\alpha_s^2)\) )  for non-vanishing $\bar \theta$, confirming that the factorization theorem captures
the leading singular behavior of the OPCC. A non-trivial check occurs at \({\cal O}(\alpha_s^2)\), where in full QCD calculation, the soft gluon splitting \(g\to q\bar q\) could in principle generate an additional charge-sensitive singular
contribution. For the net OPCC, the potentially dangerous singular term is odd under
\(q\leftrightarrow\bar q\), while the QCD matrix element and unresolved phase
space are symmetric under \(q\leftrightarrow\bar q\). Its singular contribution therefore cancels after the charge-conjugate integration. This cancellation is essential for establishing the IRC finiteness of the OPCC.

\textbf{\emph{{\color{magenta}Predictions.}}}
 We present predictions for the unpolarized OPCC distribution and for the Sivers asymmetry for EIC kinematics. 
\begin{figure}[!htbp]
    \includegraphics[width=0.5\textwidth]{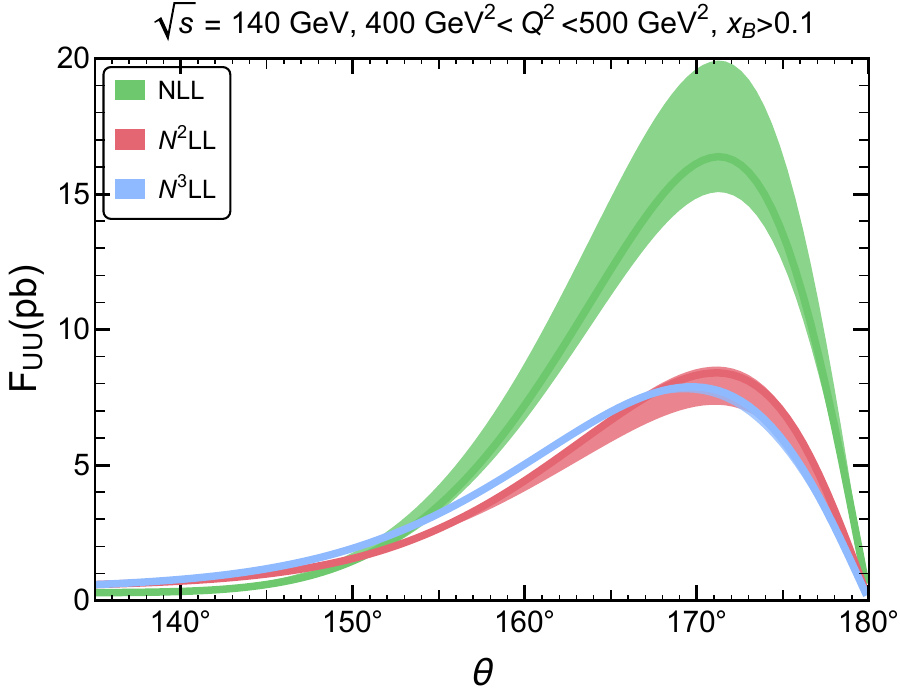}
    \caption{Comparison of NLL, N$^2$LL and N$^3$LL resummed OPCC distributions in $F_{UU}$ at the EIC kinematics. }
    \label{fg:cfrpythia}
\end{figure}
Fig.~\ref{fg:cfrpythia} shows the resummed predictions for $F_{UU}$ at different logarithmic orders. For resummation, we evolve the hard function, the jet and TMD distributions, and the soft function from their boundaries at $\mu_H = Q$, $\mu_J = \frac{2e^{-\gamma_E}}{b^\ast}$ and $\nu_J = Q$, and $\mu_S = \nu_S = \frac{2e^{-\gamma_E}}{b^\ast} $, respectively, to $\mu = \nu = Q$. We adopt the $b \to b^\ast = b/\sqrt{1+b^2/b_{\rm max}^2}$ prescription to avoid the Landau pole. We parameterize the TMD non-perturbative evolution kernel as $\exp\left(-\frac{g_2}{2}\ln\frac{Q}{Q_0}\ln\frac{b}{b^*}-g_1 b^2\right)$ with 
$g_2 = 0.84\,,Q_0 = 1.55 \, \text{GeV}\,, b_{\rm max} = 1.5{\rm GeV}^{-1}$ and $g_1 = 0.106$ and $g_1 = 0.14$ for the unpolarized TMD distribution and the jet function, respectively~\cite{Li:2021txc,Cao:2023qat,Cao:unpol,Gao:2025cwy}. 
We vary $\mu$ and $\nu$ by a factor of $2$ to estimate the theoretical uncertainty. From Fig.~\ref{fg:cfrpythia}, we see that the perturbative series stabilizes as the logarithmic accuracy is increased, and the scale uncertainty is correspondingly reduced.

  We next predict the OPCC Sivers asymmetry defined as $A^{\rm Sivers} = F_{UT}/F_{UU}$. We implement the matching in Eq.~\eqref{eq:SivmQS} at the TMD distribution boundary $\mu_J = \frac{2e^{-\gamma_E}}{b^\ast},\nu_J=Q$ with the NLO matching coefficient~\cite{Scimemi:2019gge}
 \begin{align}
& C_{f\leftarrow i}\left(x_1,x_2,b;\mu_J,\nu_J\right)   \nonumber \\ 
= &   \delta_{fi}\, \left(1-\frac{\alpha_s C_F}{2 \pi } \frac{\pi^2}{12}\right)  \delta(1-x_1)\, \delta(1-x_2) \nonumber \\ 
 &  + \delta_{fi} 
 \frac{\alpha_s}{2\pi} 
\left(C_F - \frac{C_A}{2} \right)(1-x_1) \delta\left(1-\frac{x_2}{x_1}\right)
\,.
 \end{align}
 We follow Ref.~\cite{Echevarria:2014xaa, Echevarria:2020hpy} to parameterize the Qiu-Sterman function at $\mu_0 = \sqrt{1.9}$ GeV as 
 \begin{align}
    T_{F\, i/p}(x,x,\mu_0) & = N_i\frac{\left( \alpha_i+\beta_i \right)^{\left( \alpha_i+\beta_i \right)}}{\alpha_i^{\alpha_i} \beta_i^{\beta_i}}
    \nonumber \\ 
 & \times
    x^{\alpha_i}(1-x)^{\beta_i}\,
f_{i/p}(x,\mu_0)\,.
\end{align}
where $f_{i/p}$ is the unpolarized collinear PDF, and $N_i$, $\alpha_i$ and $\beta_i$ are set to the values fitted in Ref.~\cite{Echevarria:2020hpy}. $T_{F\, i/p}(x,x,\mu_0)$ is evolved to $\mu_J$ using the relevant DGLAP evolution~\cite{Echevarria:2020hpy} to evaluate $f_{1T}^{\perp,f}(x,b,\mu_J,\nu_J)$. We  implement the N$^2$LL TMD resummation for $F_{UT}$ and the same Collins-Soper kernel is used as the unpolarized case, with $g_1 = 0.18$. 
\begin{figure}[!htbp]
    \includegraphics[width=0.45\textwidth]{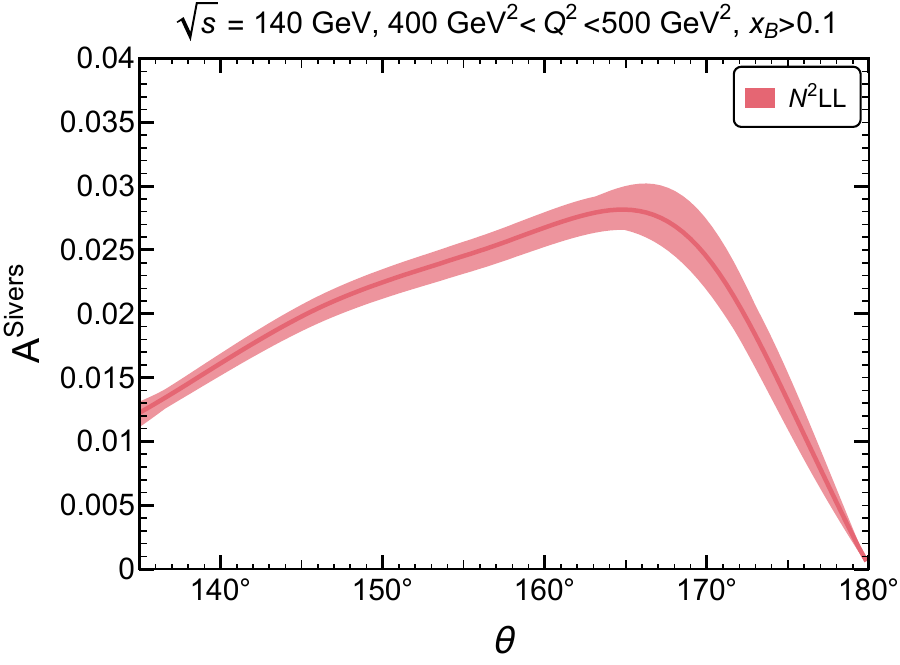}
    \caption{N$^2$LL resummed prediction for the OPCC Sivers asymmetry at the EIC kinematics. The error band indicates the scale uncertainty.}
    \label{fg:siver}
\end{figure}
Fig.~\ref{fg:siver} shows the N$^2$LL prediction for the Sivers asymmetry $A^{\rm Sivers}$ assuming EIC kinematics. We see that the OPCC probe predicts a sizeable Sivers asymmetry with systematically-improveable theoretical uncertainties.

\textbf{\emph{{\color{magenta}Conclusions and Outlook.}}} We have proposed a one-point charge correlator (OPCC) as a probe of the Sivers effect in back-to-back deep-inelastic scattering. In the TMD region, charge conservation and charge-conjugation invariance of the QCD soft sector lead to a TMD factorization theorem and render the net-charge measurement IRC finite. The measured azimuthal direction of the net charge flow can then be correlated with the transverse spin of the proton, giving a charged-track analogue of the usual Sivers modulation.

The OPCC can be predicted with systematically-improveable theoretical accuracy using currently
available perturbative ingredients, reaching N\(^3\)LL accuracy for the
unpolarized distribution and N\(^2\)LL accuracy for the Sivers asymmetry.  At the same time, the experimental measurement is exceptionally simple. It uses only the charge signs and angular directions of charged tracks, with no
final-state energy measurement, hadron identification, or jet reconstruction. This combination of theoretical control and experimental minimality makes the OPCC a clean and promising probe of the Sivers effect at a future EIC and related facilities.

While our analysis has focused on the Sivers asymmetry, the logic of the OPCC is more broadly applicable to other similar TMD observables. The same idea may also replace the reconstructed jet in lepton-jet imbalance measurements~\cite{Liu:2018trl,Liu:2020dct,Arratia:2020nxw,Kang:2020fka,H1:2021wkz,Arratia:2022oxd,H1:2024mox} by the charge-flow direction, defining a back-to-back lepton--charge-flow imbalance for TMD studies.
Analogous conserved-charge correlators can be formulated in the target-fragmentation region, in parallel with the nucleon energy correlator (NEC)~\cite{Liu:2022wop,Cao:2023oef,Liu:2024kqt,Chen:2024bpj,Zhu:2025qkx,Gao:2025cwy}. In that case, replacing energy flow by charge flow would provide a complementary charged-track probe of the angular structure of target remnants and fracture-type correlations. One application of this extension would be to revisit the light-quark dipole-operator constraints proposed with the nucleon energy correlator in Ref.~\cite{Huang:2025ljp} using a nucleon-charge correlator. 
 We thus expect the OPCC to open a new charge-and-angle-based route to nucleon structure, connecting the experimental simplicity of tracking measurements with precise theoretical control.

\begin{acknowledgments}
 \textbf{\emph{Acknowledgements.}} 
H.~C. and F.~P. are supported by the U.S. Department of Energy, Office of High Energy Physics, under contract No. DE-SC0010143. H.~C. is partially supported by a CFNS  Joint Postdoctoral Fellowship. X.~L. is supported by the National Natural Science Foundation of China under Grant No.~12547109 and Fundamental Research Funds for the Central Universities, Beijing Normal University. This research was supported in part through the computational resources and staff contributions provided for the Quest high performance computing facility at Northwestern University which is jointly supported by the Office of the Provost, the Office for Research, and Northwestern University Information Technology. 
\end{acknowledgments}




\bibliographystyle{h-physrev}   
\bibliography{ref-sivers}

\clearpage


\end{document}